\documentclass[aps,prl,floatfix,showpacs,twocolumn,superscriptaddress]{revtex4}

\usepackage{amsmath}
\usepackage{graphicx}

\newcommand{\bra}[1]{\mbox{$\langle #1|$}}
\newcommand{\ket}[1]{\mbox{$|#1\rangle$}}

\newcommand{\Ket}[2]{|#1\rangle^{(#2)}}

\begin{document}

\title{Efficient Parity Encoded Optical Quantum Computing}

\author{Alexei~Gilchrist}
\email{alexei@physics.uq.edu.au}
\affiliation{Centre for Quantum
Computer Technology and Physics Department, University of Queensland, QLD 4072, Brisbane,
Australia.}

\author{A.~J.~F.~Hayes}
\affiliation{Centre for Quantum
Computer Technology and Physics Department, University of Queensland, QLD 4072, Brisbane,
Australia.}

\author{T.~C.~Ralph}
\affiliation{Centre for Quantum
Computer Technology and Physics Department, University of Queensland, QLD 4072, Brisbane,
Australia.}

\date{\today}

\begin{abstract}
	We present a linear optics quantum computation scheme with a
	greatly reduced cost in resources compared to KLM. The scheme makes
	use of elements from cluster state computation and achieves
	comparable resource usage to those schemes while retaining the
	circuit based approach of KLM.
\end{abstract}


\maketitle


	There are a number of proposed physical systems for implementing
	quantum computation, and it is not yet clear which architecture
	would be most suitable. For initial steps toward quantum
	computation, optical systems have some appealing properties. The
	qubits are subject to low decoherence and single qubit unitaries
	can be implemented with passive linear elements. 
	
	The optical proposal by Knill, Laflamme and Milburn (KLM) \cite{klm}
	attracted much attention as it demonstrated that scalable linear-optics
	quantum computation (LOQC) was possible. KLM's proposal replaces the
	normally required large nonlinearities with nondeterministic (but heralded)
	schemes. They show that the nondeterminism can be effectively hidden by
	using a combination of teleportation (also nondeterministic) and error
	encoding. Although they showed that LOQC was possible, the resources
	consumed by their scheme are large.  Given that the overall leanness in
	consuming resources will be one of the deciding factors in adopting a
	particular implementation \footnote{Of course, there is some ambiguity in
	specifying what are to be the resources. For instance, whether we count
	single photon states consumed or Bell states will depend on the nature of
	the sources that are developed. In this paper, we shall follow the example
	of \cite{browne} and count the number of Bell states consumed as the
	primary resource in our scheme.}, the longer term prospects for optical
	schemes did not appear so great.
	
  	This changed with the alternative approach to LOQC proposed by Nielsen
	\cite{04nielsen}. This approach combined the model of cluster-state
	quantum computation \cite{briegel} with the non-deterministic gates
	presented by KLM. Cluster state computation divides the computation
	into two stages --- firstly, preparing a massively entangled state
	(the cluster state), and secondly performing the computation by a
	series of measurements on the cluster components. In Nielsen's
	scheme the cluster state preparation is non-deterministic. Once the
	cluster state is prepared, the computation proceeds
	deterministically requiring only single qubit operations and
	measurements with feed-forward. For a related scheme see also the
	approach by Yoran and Resnik \cite{yor03}.

	Creating the cluster uses much fewer resources than the KLM
	proposal. Recently, a modified scheme for preparing an optical
	cluster was proposed which uses fewer resources still and is the
	most efficient implementation yet \cite{browne}.

	In this paper we present a method which combines the ideas in both
	approaches (KLM and cluster-state). We will use the teleported
	gates of KLM with the qubits especially encoded to protect from
	teleporter failures and computational basis measurements. To build up the
	encoding and perform gates on the encoded qubits we will use the
	incremental approach in
	\cite{04hayes}. The basic encoding gate is the same
	as the type-II fusion gate of \cite{browne} and the gate we
	will need for building the resource for teleportation is
	essentially the type-I gate. The main difference is that our
	encoding is like a mini-cluster linked together with \textsc{cnot}
	instead of a \textsc{csign}.

	The motivation for this synthesis is that it retains the standard circuit
	model whilst attaining a similar reduction in resources as the cluster
	state approach. An efficient encoding against loss errors is also known for
	parity states \cite{05ralph}.

	\textbf{Notation:} since we will deal with qubits at different
	levels of encoding some care needs to be taken to establish a clear
	notation. At the highest level, the \emph{logical qubits} are
	encoded across many \emph{physical qubits} in some encoding. 
	We shall use the notation $\Ket{\psi}{n}$ to mean the
	logical state $\ket{\psi}$ of a qubit, which is encoded across $n$
	physical qubits.	The basic \emph{physical qubits} are the first
	level, and we will often drop the superscript for this level. 

	\textbf{Encoding the qubits.} The basic physical states we will use
	to construct the qubits will be the polarisation states of a photon
	so that $\Ket{0}{1}\equiv\ket{H}$ and $\Ket{1}{1}\equiv\ket{V}$. In fact, any qubits
	formed in a ``dual-rail'' fashion, i.e. by the occupation of one of two
	orthogonal modes, will do equally well and our results below can be easily
	cast in dual-rail form if desired. The advantage of this choice in optics,
	is that we can perform any single physical-qubit unitary
	\emph{deterministically} with passive linear optical elements.  It
	is interesting to note that with this choice unitary transformations will  conserve
	energy (photon number).
	
	The particular encoding we will use will be the even and odd parity
	states so that 
	\begin{eqnarray}
	\label{parity}
	\Ket{0}{n} & \equiv & (\ket{+}^{\otimes n}+\ket{-}^{\otimes
	n})/\sqrt{2}\nonumber \\  
	\Ket{1}{n} & \equiv & (\ket{+}^{\otimes n}-\ket{-}^{\otimes
	n})/\sqrt{2},
	\end{eqnarray}
	where  $\ket{\pm} = (\ket{0} \pm \ket{1})/\sqrt{2}$.	
	A useful feature to note is that the parity states can be written as any
	sum where each term has the original parity e.g.
	$\Ket{0}{n}=(\Ket{0}{n-1}\ket{0}+\Ket{1}{n-1}\ket{1})/\sqrt{2}$.
	This choice of encoding means that a computational
	basis measurement of one of the physical qubits will not destroy
	the logical state, but will only reduce the level of encoding.  To see this
	notice from Eq.\ref{parity} that $\langle 0|0\rangle^{(n)} = \Ket{0}{n-1}$
	and $\langle 0|1\rangle^{(n)} = \Ket{1}{n-1}$, and thus $\langle
	0|\psi\rangle^{(n)}= \Ket{\psi}{n-1}$. On the other hand $\langle
	1|0\rangle^{(n)} = \Ket{1}{n -1}$ and $\langle 1|1\rangle^{(n)}=
	\Ket{0}{n-1}$, and in this case a bit flip of the logical qubit has
	occurred. However, this can be 	easily corrected because a bit flip of any
	one physical qubit will bit flip the logical qubit. Thus $X \langle
	1|\psi\rangle^{(n)} = \Ket{\psi}{n-1}$ where
	$X=\ket{0}\bra{1}+\ket{1}\bra{0}$ is the usual Pauli-$X$ operator.	

\begin{figure}[htpb]
	  \begin{center}
	\begin{tabular}{cccc}
	(a) & \includegraphics{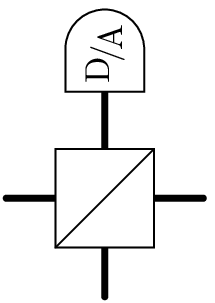} & (b) & \includegraphics{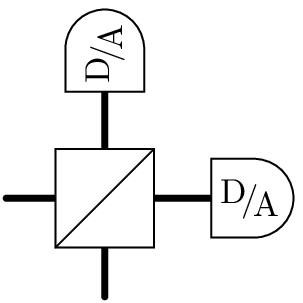}\\
	(c) & \multicolumn{3}{l}{\includegraphics{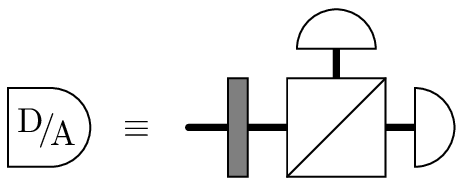}} \\
	(d) & \includegraphics{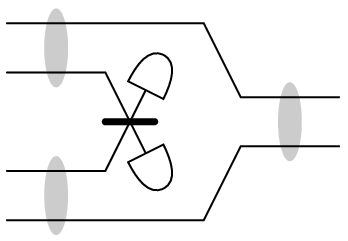} & (e) & \includegraphics{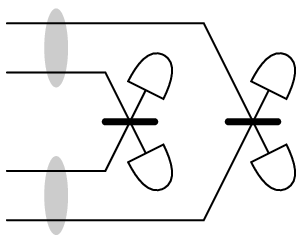} 
	\end{tabular}
	  \end{center}
		\caption{(a) The type-I fusion ($f_I$) and (b) type-II fusion ($f_{II}$) gates. 
		(c) a detector analysing in the 
		diagonal-antidiagonal basis. Dual-rail forms of the $f_I$ and $f_{II}$ gates are shown 
		in (d) and (e) respectively (the shaded ellipses represent the dual-rail qubits.}
	  \label{fig:fusion}
	\end{figure}

	The key functional components in our scheme are two gates which we
	will call type-I ($f_I$) and type-II ($f_{II}$) fusion gates
	following the nomenclature of \cite{browne}. These gates are shown
	in Fig.~\ref{fig:fusion} as polarisation and dual-rail gates. 
	The action of the gates can be represented in short-hand as POVM measurement
	operators with the result being particular detector states denoted
	as $d_{1010}$ for the detector sequence ``1,0,1,0'' etc. With this notation,
	the sucessful $f_{II}$ operators are
	\begin{align}
	\ket{d_{1010}}(\bra{00}+\bra{11}), \ket{d_{0101}}(\bra{00}+\bra{11})\\
	\ket{d_{1001}}(\bra{00}-\bra{11}), \ket{d_{0110}}(\bra{00}-\bra{11})
	\label{eq:fIIPOVMs}
	\end{align}
	and the unsucessful ones are 
	\begin{align}
	\ket{d_{2000}}\bra{01}, 
	\ket{d_{0200}}\bra{01},
	\ket{d_{0020}}\bra{10},
	\ket{d_{0002}}\bra{10}.
	\label{eq:fIIPOVMf}
	\end{align}
	Note that even without photon-number discriminating detectors these events are
	distinguishable. This is not true for $f_I$ events which have the following successful
	operators
	\begin{align}
	\ket{d_{10}}(\ket{0}\bra{00}+\ket{1}\bra{11}),\\ 
	\ket{d_{01}}(\ket{1}\bra{00}-\ket{1}\bra{11})
	\label{eq:fIPOVMs}
	\end{align}
	and unsuccessful operators
	\begin{align}
	\ket{d_{20}}\ket{\dots}\bra{01}, 
	\ket{d_{02}}\ket{\dots}\bra{01},
	\ket{d_{00}}\ket{\dots}\bra{10}
	\end{align}
	which all measure in the computational basis and project the remaining mode
	outside of that basis.

    The fusion gates act as partial Bell measurements on the input
    qubits, and are used to implement entangling operations. Such
    non-deterministic Bell measurements have been essential in
    attempting to use linear optics for quantum communication and
    computation. In 1994 Weinfurter \cite{wein94} described an optical layout that
	used beam-splitters and detectors to distinguish two of the four
	Bell states on spatially-encoded qubits. It was observed that this
	configuration could be used to teleport qubits with a success
    probability of 50\%. Soon after, Braunstein and Mann \cite{braun95}
	published a similar scheme that acted on polarisation-encoded
	qubits. The optical configuration they provide is equivalent to
	the $f_{II}$ gate. In both papers, what we have referred to as a
	dual-rail Bell measurement was used, measuring both states that
    the photon could occupy. Calsamiglia and L\"utkenhaus \cite{casnlut00}
    later demonstrated that this 50\% probability of uniquely distinguishing
	a Bell state was the best that could be achieved using linear
    optical components. However, as shown in KLM \cite{klm}, the probability of
	successfully teleporting a qubit can be made arbitrarily high if
	sufficient resources are used.

	The type-II fusion gate can be used to add $n$ physical
	qubits to the encoded state by using a resource of
	$\Ket{0}{n+2}$. When successful (with probability $1/2$),
	then the following takes place (with a bit-flip applied in
	half the cases):
	\begin{equation}
		f_{II}\Ket{\psi}{m}\Ket{0}{n+2}\rightarrow\left\{ \begin{array}{cl}
			\Ket{\psi}{m+n} & \mbox{(success)}\\
			\Ket{\psi}{m-1}\Ket{0}{n+1} & \mbox{(failure)}
		\end{array}\right.
		\label{eq:encoding}
	\end{equation}
	
	If the gate fails then a physical qubit is removed from the
	encoded state, and the resource state is left in the state
	$\Ket{0}{n+1}$ which can be recycled. It will again be
	necessary to apply a bit-flip correction in half the cases. 
	This encoding procedure is equivalent to a gambling game
	where we either lose one level of encoding, or gain $n$
	depending on the toss of a coin.

	\textbf{Generating the resource.}
	Given a supply of Bell states ($\Ket{0}{2}$), the resource
	$\Ket{0}{n}$ can be built up using the same techniques as used to
	build up cluster states given in \cite{browne}. In fact,
	$\Ket{0}{n}$ is nothing more than a linear graph state built up
	with \textsc{cnot}s, instead of \textsc{csign} gates as for the
	cluster states.
	
	To create the state $\Ket{0}{3}$, two $\Ket{0}{2}$ can be fused
	together using the $f_I$ gate. When successful, the $\Ket{0}{3}$
	state is produced, when unsuccessful, both bell states are
	destroyed. Since $f_I$ functions with a probability of $1/2$, on
	average two attempts are necessary so on average each $\Ket{0}{3}$
	consumes $4 \Ket{0}{2}$. 
	
	Once there is a supply of $\Ket{0}{3}$ states, either $f_I$ or
	$f_{II}$ can be used to further build up the resource state via 
	\begin{align}
		H f_I(H\!\otimes\! H)   \Ket{0}{n}\Ket{0}{m}&\rightarrow \left\{ \begin{array}{cl}
			\Ket{0}{m+n-1} & \mbox{(success)}\\
			- & \mbox{(failure)}
		\end{array}\right.\label{eq:fIjoin}\\	
		f_{II} \Ket{0}{n}\Ket{0}{m}&\rightarrow \left\{ \begin{array}{cl}
			\Ket{0}{m+n-2} & \mbox{(success)}\\
			\Ket{0}{m-1}\Ket{0}{n-1} & \mbox{(failure)}
		\end{array}\right.	
		\label{eq:fIIjoin}
	\end{align}
	In the first case we use $f_I$ with Hadamard gates and this
	approach has the advantage of losing only a single qubit from the
	input states, but the disadvantage of completely destroying the
	entanglement in both input states in the event of failure. In the
	second case, we use $f_{II}$ to join the input states at the
	expense of losing two of the initial qubits. There are
	two advantages to the second scheme --- in the case of failure we
	do not destroy the entanglement of the input states, just reduce
	their encoding by one; and we do not need photon number
	discriminating detectors to operate $f_{II}$.

	Despite the advantages in using $f_{II}$, numerical exploration
	seems to indicate that simply fusing two $\Ket{0}{3}$ with $f_I$
	to form a $\Ket{0}{5}$ is near optimal. This approach carries an
	average cost of $16 \Ket{0}{2}$ per $\Ket{0}{5}$. Only once we have
	a supply of $\Ket{0}{5}$ is it advantageous to switch to another
	strategy using $f_{II}$, and incrementally add $\Ket{0}{5}$ to the
	resource.

	\textbf{Gates on the logical states.}
	With the parity encoding we can deterministically perform any of the
	gates that can be achieved with the set $\{X_\theta, Z\}$ on a
	logical qubit. Here the notation is $X_\theta = \cos(\theta/2)I+i\sin(\theta/2)X$. 
	The $Z$ gate on a logical qubit can be performed by
	applying a $Z$ gate on all the physical qubits. Since the number of
	sign flips obtained in this way will depend on the parity of the
	state, this will have the desired effect on the logical state. To
	perform an arbitrary $X$ rotation on a logical qubit, we can apply
	that rotation to any of the physical qubits.

	In order to get a universal set of gates we need to also perform
	the set $\{Z_{90}, CNOT\}$. These gates need to be performed
	non-deterministically on the encoded qubits, and performing these
	gates efficiently is the principal aim of this paper.


		The non-deterministic single logical-qubit gate we need is a
		straightforward extension of the encoding procedure. We simply perform
		a $Z_{90}$ gate on one of the physical qubits and re-encode from that
		qubit. As before, if the re-encoding is unsuccessful we lose an
		encoding level from the logical qubit (the effect of the $Z_{90}$
		appears as a global phase shift in this case). If the encoding is
		successful on the other hand, we can now measure the remainder of the
		physical qubits in the computational basis and if the parity of the
		result is odd, apply a bit flip for correction. Note that it would be
		possible to perform a $Z_{\theta}$ in the same way but the parity
		measurement would randomly flip the angle to $-\theta$ half the time so
		this is not so useful.

		Creating a \textsc{cnot} proceeds along very similar lines to the
		$Z_{90}$ gate. We first entangle the control qubit with one qubit of an
		$\Ket{0}{m+1}$ resource with an $f_I$ fusion gate. We then entangle one
		component of the target qubit with the output of the above operation
		using an $f_{II}$ fusion gate. At this stage, if we were to measure the
		remaining physical qubits from the control qubit and apply a bit-flip
		depending on the parity of the result, we would have performed the
		operation:
		\begin{equation}
			\Ket{\psi}{n,n}\rightarrow \mbox{\textsc{cnot}}\Ket{\psi}{m,n-2}
			\label{eq:cnot}
		\end{equation}

		\begin{figure}[htpb]
		  \begin{center}
		\begin{tabular}{cc}
		(a) & \includegraphics[scale=.8]{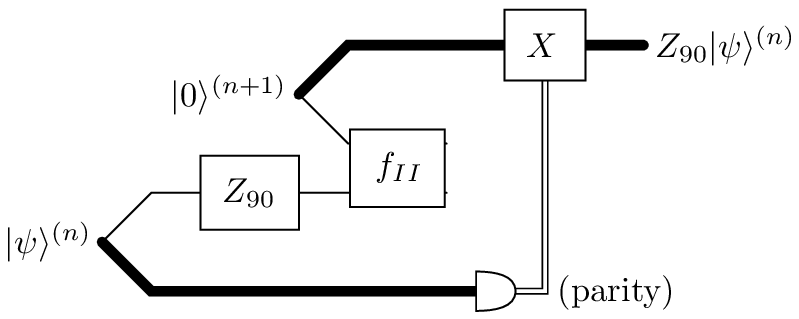} \\
		(b) & \includegraphics[scale=.8]{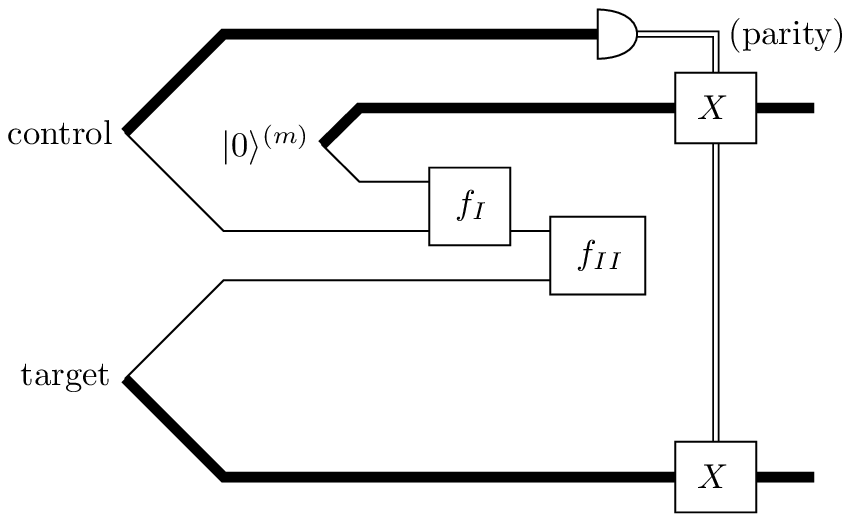} 
		\end{tabular}
		  \end{center}
			\caption{(a) Implementation of $Z_{90}$ gate and (b) \textsc{cnot}
			gate. Note that the \textsc{cnot} gate is symmetric --- by choosing
			to measure the parity of the lower qubit instead, the roles of
			target and control will be swapped.}
		  \label{fig:gates}
		\end{figure}
	
		Instead of immediately re-encoding the target or control
		qubits to the full amount, a non-deterministic
		$Z_{90}$ operation can be incorporated into the
		process, thereby effectively getting it for free.
		
		We can employ recycling of entangled states in our scheme as was
		done for the cluster state proposal of \cite{browne}.  The
		failure modes of the $f_{I}$ and $f_{II}$ fusion gates do not destroy the
		resource but reduce its encoding by one. This resource state can
		then be recycled for further attempts. 		It should be noted
		however that recycling may not be particularly useful in either
		scheme. The optics and control necessary to recycle these states
		without degradation will probably be more difficult than simply
		producing more resource states from scratch.

	\textbf{Probability and resources.}
	What level of encoding do we need to maintain? If we choose too low
	a level then there is a significant probability that we can destroy
	the logical qubit through a long string of failures. If the
	encoding is too high then this carries an unnecessarily high
	resource usage.

	Consider performing a $Z_{90}$ gate on a qubit encoded across
	$n$ physical qubits. Since the fusion gates fail with probability
	$1/2$, and assuming we will re-encode to the full amount $n$ in one
	hit, then the success probability of the gate will be $1-(1/2)^n$.
	The average resource requirements will just be twice the resource
	requirements needed to generate the state $\Ket{0}{n+1}$. These figures
	are shown in table~\ref{tab:probs} (b).

	If we are re-encoding in multiple smaller steps then this figure will be an
	\emph{upper} bound on the success probability. The advantage of re-encoding
	in smaller steps is that we can consume fewer resorses. Probabilities and
	resources for an alternative re-encoding strategy is shown in
	table~\ref{tab:probs} (c).

	\begin{table}[htb]
		\begin{center}
			\begin{tabular}{c|ccccc}
				$m$ & (a) & (b) & (c) & (d) & (e) \\
				\hline
				3  & 4  & 4  & 87.48\% & 16  & 19 \\
				4  & 10 & 10 & 93.76\% & 28  & 25 \\
				5  & 16 & 16 & 96.96\% & 51  & 45 \\
				6  & 28 & 28 & 98.47\% & 76  & 53 \\
				7  & 40 & 38 & 99.20\% & 101 & 63 \\
				8  & 52 & 44 & 99.58\% & 126 & 78 \\
				9  & 64 & 57 & 99.80\% & 174 & 90 \\
				10 & 88 & 66 & 99.89\% & 222 & 100  
			\end{tabular}
		\end{center}
		\caption{Sucess probabilities and resource usage. (a) Average
		number of Bell states consumed in forming $\Ket{0}{m}$ using
		$f_I$ and no recycling. (b) Average number of Bell states consumed
		with recycling when advantageous.
		(c) Probability and (d) average resource
		usage of sucessfully performing $Z_{90}$ and re-encoding in one
		step. The resource usage is simply the cost of generating
		$2\Ket{0}{n+1}$, and involves no recycling. (e) 
		resource usage to performg $Z_{90}$ using
		recycling. Values
		calculated numerically from 500,000 runs.}
		\label{tab:probs}
	\end{table}

	For the \textsc{cnot} gate depicted in figure~\ref{fig:gates}(b), when the
	encoding is sufficient that there are no boundary effects, the success
	probability is simply $1-(3/4)^n$  since both the $f_I$ and $f_{II}$ gates
	have to succeed. At the conclusion of the gate the control qubit is left
	encoded across $m$ physical qubits (assuming a resource $\Ket{0}{m+1}$),
	and the target will on average lose two physical qubits from the encoding.

	The success probability can be boosted by first pre-encoding the top qubit
	of figure~\ref{fig:gates}(b) to boost the size of the parity code. After
	the gate is successful, the measurement of parity can be delayed and the
	size of the top qubit can again be increased by appending some more
	resource.  The results of a strategy implementing this are shown in
	table~\ref{tab:probs-cnot}.

	\begin{table}[htb]
		\begin{center}
			\begin{tabular}{c|ccc}
				$n$ & (a) & (b) & (c) \\
				\hline
				6  & 96.4\% & 181 & 115 \\
				7  & 97.6\% & 190 & 117 \\
				8  & 98.2\% & 196 & 121 \\
				9  & 98.6\% & 208 & 126 \\
				10 & 98.9\% & 228 & 151  
			\end{tabular}
		\end{center}
		\caption{Sucess probabilities and resource usage for a \textsc{cnot}.
		Both qubits initially are encoded across $n$ physical qubits.  The
		strategy involved pre-encoding the control (with resource size
		$\Ket{0}{8}$ for all except $n=6$ where it was $\Ket{0}{7}$) if it fell below 6
		pariy qubits; using $\Ket{0}{5}$ in the actual gate itself; and
		post-encoding back up to the initial $n$ encoding level once the gate
		was successful.  (a) Probability of operation. Average resources
		consumed with (b) no recycling and (c) with recycling are also shown.
		Values calculated numerically from 100,000 runs.}
		\label{tab:probs-cnot}
	\end{table}

\textbf{Conclusion}
In this paper we have presented a method for KLM-style optical
quantum computation which dramatically reduces the resource usage
over the original scheme. The method compares favourably against the
most resource-efficient cluster state schemes, achieving comparable
resource usage. In fact, our method combines techniques from both
approaches. We borrow the circuit based approach and parity encoding
from the KLM proposal, and the method of resource preparation from the
cluster state approach.


We would like to acknowledge helpful discussions with Bill Munro and 
Stefan Scheel.


\end{document}